\documentstyle[graphicx,multicol,prl,aps]{revtex}

\newcommand {\be} {\begin{equation}}
\newcommand {\bea} {\begin{eqnarray} \nonumber }
\newcommand {\ee} {\end{equation}}
\newcommand {\eea} {\end{eqnarray}}

\begin{document}
\parskip=0cm
\noindent

\title {On the Effects of a Bulk Perturbation on the
Ground State of 3D Ising Spin Glasses}

\author{Enzo Marinari and Giorgio Parisi}

\address{
  Dipartimento di Fisica, INFM and INFN, 
  Universit\`a di Roma {\em La Sapienza},\\
  P. A. Moro 2, 00185 Roma (Italy).
}

\date{\today}                                                 

\maketitle

\begin{abstract}  
We compute and analyze couples of ground states of $3D$ spin glasses
before and after applying a volume perturbation which adds to the
Hamiltonian a repulsion from the true ground state.  The physical
picture based on Replica Symmetry Breaking is in excellent agreement
with the observed behavior.
\end{abstract}


\begin{multicols}{2}
\narrowtext
\parskip=0cm

Studying the scaling behavior of ground states (GS) of disordered
systems \cite{GSONE} turns out to be a powerful tool. Here by using a
powerful technique based on a bulk modification of the Hamiltonian
\cite{PALYOU2,MPRRZ} we are able to give strong numerical evidence of
a Replica Symmetry Breaking like behavior of $3D$ spin glasses. With
this study we believe we vastly enlarge the scope of \cite{PAPONE},
where we were changing the boundary conditions to analyze the changes
induced in the GS microscopic structure.  The open debate is about the
role of the exact ({\em Replica Symmetry Breaking}, RSB) solution of
the Mean Field spin glass model, where field theory is exact by
definition \cite{RSB}, in describing finite dimensional spin glasses:
the validity of a possible phenomenological description, the (more or
less modified) {\em droplet picture} \cite{DROPLET}, should probably
be taken as an alternative possibility.  The RSB picture is
characterized by the presence of states (for the terminology we refer
to \cite{MPRRZ}) that are locally different one from the other.

In this respect equilibrium Monte Carlo simulations are useful
\cite{MPRRZ}, but since it is impossible to thermalize large, low $T$
systems, if one is not far enough from $T_c$ there is the remote
possibility \cite{BRAMO} that the asymptotic physical picture could be
disguised because of crossover effects (this criticism does not apply
to the off-equilibrium simulations \cite{RIRI}, and very low $T$
recent simulations \cite{KAPAYO} also claim to disprove it).  The
remarkable achievement of the work of \cite{GSONE} has been to realize
that one can compute GS for systems of size comparable to the one of
finite $T$ numerical simulations.  The first series of papers
\cite{GSONE,PAPONE} has been computing couples of GS with periodic
boundary conditions (PBC) and with anti-periodic boundary conditions
(APBC).  A different method, based on the use of a bulk perturbation
\cite{PALYOU2} has been first used by Palassini and Young to detect
the inconsistency of a (even modified) droplet picture: large scale
excitations with a finite energy cost (first discussed in
\cite{MARTIN}) are clearly detected by this approach.  In the replica
approach we expect that these large scale excitations correspond to
the simultaneous reverse of a large non-compact block of spins: the
surface to volume ratio of the set of the flipped spins should
asymptotically go to a non-zero value.  In previous studies
\cite{MARTIN,PALYOU2} this surface to volume ratio was investigated:
it was found that it was slowly decreasing with $L$ and its behavior
was compatible with a small negative power of $L$.

We use here the same method to make stronger the evidence of the
existence of these large scale excitations, and we present a detailed
quantitative study of their properties.  We show that they behave in
agreement with RSB, and that the surface to volume ratio extrapolates
to a non-zero value in the limit $L \to \infty$.

We work in three spatial dimensions ($D=3$) on simple cubic lattices
of linear size $L$, containing $N=L^3$ spins with PBC (we have taken
$L=4$, $6$, $8$, $10$, $12$, $14$).  We consider quenched random
couplings $J$ extracted from a Gaussian probability distribution 
with zero expectation value and unit variance.  We study the behavior
of the GS after changing the Hamiltonian to include a term which
repels from the true GS.  The Hamiltonian of the first system is

\begin{displaymath}
  H_{0}(\{\sigma\})\equiv 
  -\sum_{i,\mu} \sigma(i) J(i,\mu) \sigma(i+\hat\mu)\ ,
\end{displaymath}
where the sum is over the $N$ sites $i$ and the $D$ directions $\mu$
($\hat\mu$ is the versor in the $\mu$ direction).  The Hamiltonian of
the second system is

\begin{displaymath}
  H_{1}(\{\tau\}) \equiv H_0(\{\tau\})
  - \frac{\epsilon}{ND} \sum_{i,\mu}
  \sigma_0(i)\sigma_0(i+\hat\mu)
  \tau(i)\tau(i+\hat\mu) \ ,
\end{displaymath}
where the $\sigma_0(i)$ are the spins in the GS of $H_{0}$.  Let
$\tau_0(i)$ be the spins in the GS of $H_{1}$: we are interested in
studying the differences among $\sigma_0$ and $\tau_0$ when $N$ goes
to infinity at fixed $\epsilon$ (in this limit the perturbation is of
order 1, while the total Hamiltonian is of order $N$).  At this end we
define the {\em local site overlap} on site $i$ as $q(i) \equiv
\sigma_{0}(i)\ \tau_{0}(i)$, and the {\em local link overlap} is
defined among nearest neighbor sites (or equivalently on links) as
$q_{l}(i,\mu) \equiv q(i)\ q(i+\mu)$.  If two spin configurations
differ by a global spin flip their link overlap is identically equal
to $+1$ (while their overlap is equal to $-1$).  We also define the
average over the system sites and links of the previous quantities:
$q\equiv N^{-1}\sum_{i}q(i)$, $q_{l} \equiv (DN)^{-1} \sum_{i,\mu}
q_{l}(i,\mu)$.  With these definitions we have that $H_{1}(\{\tau\}) =
H_0(\{\tau\}) -\epsilon q_{l}$.  We are considering this type of
perturbation because the quantity $q_{l}$ is sensitive only to local
changes of the configuration, i.e. it is different from 1 only on the
interface among the clusters of flipped and non-flipped spins.

The main effect of the perturbation is to generate excited states.  It
is reasonable to assume that for $L\to\infty$ the value of the
observables computed for fixed value of the site overlap are weakly
dependent on the value of $\epsilon$.  This is also suggested by the
results of \cite{PALYOU2}, and will be discussed in more details in
\cite{MAPA}.  For this reason we will consider the link overlap (and
also other quantities) as a function of the (square) site overlap,
i.e. $q_l(q^2)$. This quantity is defined as the average of the link
overlap over those systems for which the square site overlaps of the
two ground state is $q^2$. We will indicate the link overlap summed
over all the systems by $\overline{q}_l\equiv \int dq^{2} q_l(q^2)
P(q^{2})$, where $P(q^{2})$ is the probability distribution of
$q^{2}$.  We have computed $10^4$ GS at $L=4$ and $6$, $2000$ at
$L=8$, $1500$ at $L=10$, $500$ at $L=12$ and $14$, with
$\epsilon=4\sqrt{6}$, using a genetic algorithm which will be
described in \cite{MAPA}.

\begin{figure}
\centering
\includegraphics[width=0.45\textwidth,angle=0]{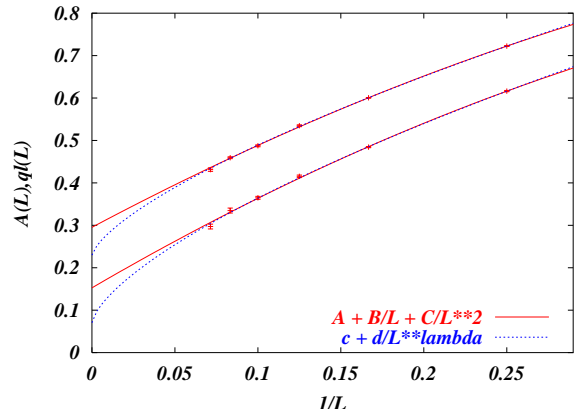}
\caption[a]{The value of $A(L)$ (upper points) and of $\overline{q}_l$
as a function of $L^{-1}$. 
The curves are for a power fit and a polynomial fit.
\protect\label{F-AQL}}
\end{figure}

Given the importance of the issue, we start by discussing the
dependence of $q_{l}(q^2)$ over $q$ and $L$.  If we plot $q_l(q^2)$
versus $q^2$, we notice that $q_l(q^2)$ and $q^2$ are very strongly
correlated. In order to clarify this effect we define the quantity
(for all systems such that $q\ne 1$) $ R_L(q^2)\equiv
(1-q_{l}(q^2))/(1-q^{2})$.  For fixed $L$ $R_L(q^2)$ is nearly $q^{2}$
independent in a large region where $q$ is not too large (only close
to $q=1$ the dependence becomes slightly more noticeable).  The
$q^{2}$ dependence of $R(q)$ in this large $q$ region becomes milder
with increasing $L$.  We fit $R_L(q^2)$ to the form $A(L)+B(L)\ q^2$
using the data in the range $q^2 \in (0,0.3)$: the fits are very good,
and the extrapolations are all very smooth.  $B(L)$ is in all cases a
few percent of $A(L)$.  In fig. \ref{F-AQL} the upper points are for
$A(L)\equiv R_L(0)=1-q_l(0)$ versus $L^{-1}$.  Both a linear fit
(which extrapolates to a non zero value) and a power fit to the form
$C L^{-\alpha}$, where $\alpha$ turns out to be $0.44$, follow the
data in a qualitative way, but they are unacceptable because they lead
to a very high value of $\chi^{2}$ (i.e. to a normalized
$\chi^{2}=O(20)$).  In fig.  \ref{F-AQL} we plot a second order
polynomial fit that is very good (it has a normalized $\chi^{2}\simeq
0.6$) and leads to a value of $A(L=\infty) = 0.30\pm 0.01$.  We also
plot a fit to the form $A(L)=a+b L^{-\lambda}$, that is also good (it
has a normalized $\chi^{2}=0.3$) and leads to a value of of
$A(L=\infty)=0.24\pm 0.01$ and $\lambda=0.72\pm 0.02$. The difference
with the $0.30$ obtained by the polynomial fit gives an idea of the
size of the systematic error over the extrapolation.

The values extrapolated to $L=\infty$ are an estimate of
$1-q_l(0)$, and they have to be compared with the value $0.245\pm
0.015$ that we have estimated in \cite{PAPONE}: the agreement is good,
even if here the systematic error is large.  We can conclude that
there is evidence that $1-q_{l}(0)$ does not extrapolate to zero
in the infinite volume limit and the dependence of $q_{l}(q^2)$ is
well approximated (for $q^{2}$ not too close to $1$) by the form
$q_{l}(q^2)=1-A(1-q^{2})$.  The previous formula holds in the infinite
range SK model, with $A=1$.  The fact that in $3D$ we find a low value
($A\simeq 0.3$) suggests that $A$ goes to zero when one approaches the
lower critical dimension.

The lower points and curves of fig. \ref{F-AQL} are for the
expectation value of the link overlap $\overline{q}_l$, averaged over
all $q^2$ values.  The curves are again a polynomial and a power law
best fits, that here extrapolate to a lower value ($0.07\pm 0.04$ for
the power law and $0.155\pm 0.015$ for the polynomial fit).  The error
is larger than in $A(L)$ (of a factor close to $2$) but apart
from a shift (since $\overline{q}_l \ne q_l(0)$) the two curves
are similar.

Let us stress that $A(L)$ is important also since $q_l(q^2)$ is not
expected to depend on $\epsilon$ when $L \to \infty$.  The
$\epsilon$-perturbation allows to explore low energy states with an
energy gap of order $1$. The function $P(q^{2})$ changes with
$\epsilon$ \cite{FRAPA} and consequently also the value
$\overline{q_l}$ changes: still the functional relation that relates
$q_l$ to $q^2$ is unchanged.


It is also interesting to look at the variance squared of $R_L(q^2)$,
i.e. to $\sigma_{(R)L}^{2}(q^2)\equiv \langle R_L^{2}(q^2)\rangle
-\langle R_L(q^2)\rangle^{2}$. We proceed as for the the best fit of
$R_L(q^2)$. First we fit $\sigma_{(R)L}^{2}(q^2)$ to the form
$A_W(L)+B_W(L)\ q^2$, using the data with $q^2\le 0.3$: the
extrapolations are very good.  Also here for a given value of $L$ we
find that $\sigma_{(R)L}^{2}(q^2)$ is nearly $q^{2}$ independent apart
from the region close to $q^{2}=1$.  Also the $q^{2}$ dependence of
$\sigma_{(R)L}^{2}(q^2)$ in this region becomes weaker mild with
increasing $L$.

We plot $A_W(L)$ versus $1/L$ in fig. \ref{FIG-WIDTH}. Together with
the data we show a best fit to the form $A_W(L) \simeq C \
L^{-\delta}$, that has a low $\chi^2$, and $\delta = 1.35 \pm
0.03$. The fact that a fit to a zero asymptotic value is very good
shows that $q_l$ as a function of $q$ does not fluctuate as
$L\to\infty$: this is the same situation that we find in the mean
field RSB theory, with a more complex dependence of $q_l(q^{2})$ over
$q^2$. The asymptotically vanishing value of $A_W(L)$ explains why the
errors on $q_l(q^{2})$ are smaller that those on $\overline{q_l}$.


We discuss next the overlap-overlap correlation functions.
The difference among a droplet and a RSB scenario becomes very clear
if we consider the overlap-overlap correlation function at fixed value
of $q^{2}$ in a system of side $L$, i.e. the correlation function
obtained by selecting those systems which  have a ground state with a
fixed value of the mutual overlap:

\begin{figure}
\centering
\includegraphics[width=0.45\textwidth,angle=0]{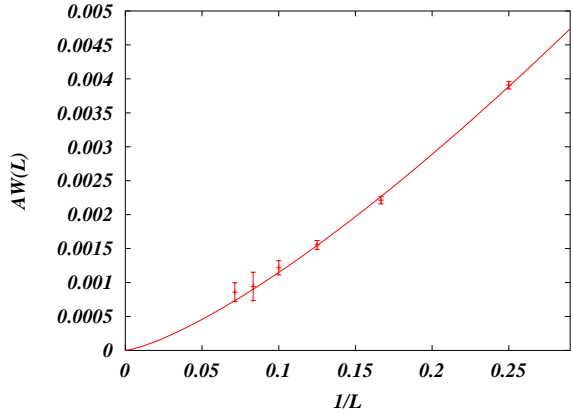}
\caption[a]{The value of $A_W(L)$ as function of $L^{-1}$. 
The curve is a power fit to a zero
asymptotic value.  \protect\label{FIG-WIDTH}}
\end{figure}

\begin{displaymath}
  C(x,L|q^{2}) \equiv \langle q(x)q(0) \rangle_{(q^{2})}\ .
\end{displaymath}
In the RSB case we expect that for large $L$ and fixed $x$
$C(x,L|q^{2}) \to G(x|q^{2}) + q^{2}$, where the function $G(x|q^{2})$
depends on $q^{2}$ and for large $x$ should behave as
$G(x|q^{2})\propto x^{-\phi(q^{2})}$. $\phi(q^{2})$ may
depend on $q^{2}$, although there are indications that it takes at
most only two different values, one at $q=0$ and one at $q\ne 0$
\cite{DEDOMI}.  In the region where $x$ and $L$ are large we expect
that

\begin{equation}
  C(x,L|q^{2}) \approx g(r|q^{2})(\frac{1}{x}+
  \frac{1}{L-x})^{\phi(q^{2})} + q^{2}\ ,
  \label{E-SCALING}
\end{equation}
where $r\equiv x L^{-1}$. We have used the factor
$(\frac{1}{x}+\frac{1}{L-x})$ instead of the simpler $\frac{1}{x}$ in
order to implement the symmetry implied
by the presence of PBC.

The situation described by a droplet picture is different: in this
case, for large $L$ and fixed $x$, $C(x,L|q^{2}) \to 1$.
One can argue that in the droplet model one should have
$\phi(q^{2})=0$ and $g(0|q^{2})+q^{2}\simeq 1$, i.e
$C(x,L|q^{2}) \approx D(xL^{-1}|q^{2})$ with $D(0|q^{2})=1$.

\begin{figure}
\centering
\includegraphics[width=0.45\textwidth,angle=0]{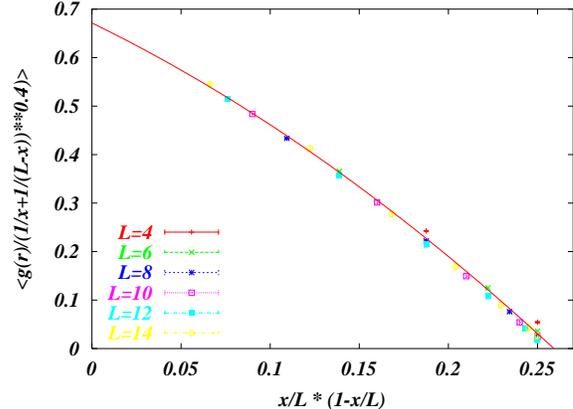}
\caption[a]{The overlap-overlap correlation function minus $q^{2}$
divided times the factor
$\left(\frac{1}{x}+\frac{1}{L-x}\right)^{0.4}$ (i.e. the function $g$
defined in (\ref{E-SCALING})) as a function of
$w=\frac{x}{L}-\left(\frac{x}{L}\right)^{2}$. The line is for the best
fit to a second order polynomial.  \protect\label{F-GR} }
\end{figure}

We will discuss our results for small values of $q$: we consider the
$q$ range $q\le q_{max}=0.1$.  The results are not very sensitive to
the exact position of the $q_{max}$ cut (of course if $q_{max}$ is too
small the statistics fades).  In fig. \ref{F-GR} we present our data
for $g(r)$ as a function of $w\equiv r(1-r)$ for $\phi=0.4$ (we prefer
to plot $g$ versus $w$ since because of the PBC both $g$ and $w$ are
left invariant when $r$ goes in $1-r$).  The value of $q^{2}$ we use
to subtract the disconnected part is the average of $q^{2}$ in the
interval $q \le q_{max}$.  We have selected $\phi$ such to optimize
the quality of the scaling behavior.  The line is for the best fit to
a second order polynomial.

The scaling is very good, and the estimate of $\phi\simeq 0.4$
reliable. A droplet-like scaling is excluded by our data.
We notice that 
$C(x,L|0)|_{x=1}= q_{l}(0)$ and the scaling form (\ref{E-SCALING})
accounts for the $L$-dependence of $q_{l}(0)$.  If we assume that
(\ref{E-SCALING}) is exact also for $x=1$ (a too strong assumption) we
get that $C(1,\infty|0)= g(0|0)$.  Using the quadratic fit in
fig. \ref{F-GR} to extrapolate the data at $r=0$ we get in the
infinite volume limit $1-q_{l}(0)=.33 \pm .02$, which is in reasonable
agreement with the previous estimates, given the crudeness of the
approximation.

\begin{figure}
\centering
\includegraphics[width=0.45\textwidth,angle=0]{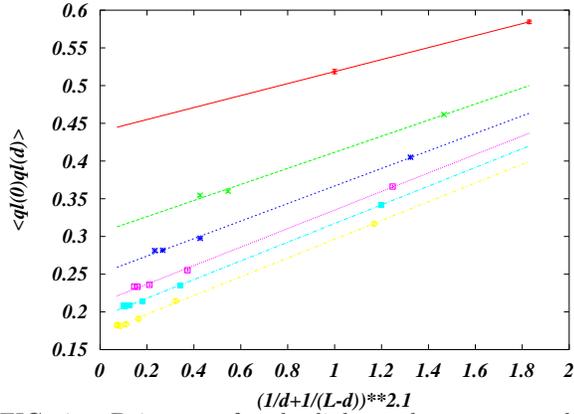}
\caption[a]{ Points are for the link overlap auto correlation
functions versus $w^{2.1}$. Lattice sizes are from $L=4$ (upper points
and line) to $L=14$ (lower points and line).  Lines are for linear
best fits.  \protect\label{F-OLOL} }
\end{figure}

\begin{figure}
\centering
\includegraphics[width=0.45\textwidth,angle=0]{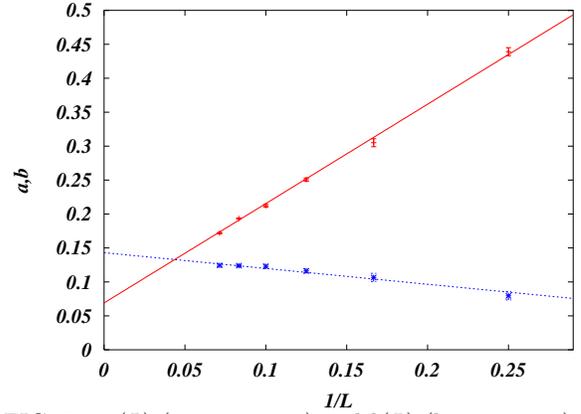}
\caption[a]{ $a(L)$ (upper points) and $b(L)$ (lower points) from
equation (\ref{E-AB}) and the two best linear fits.
\protect\label{F-AB} }
\end{figure}

When we analyze the correlation function of the link overlap we also
select events at a fixed value of $q^{2}$: as in the previous case we
will consider a cut of the form $|q|<q_{max}=.1$.  We only measure a
simple correlation function of $1-q_{l}(i)$ (we only look at
correlations in a given plane, and average the two contributions
coming from separating in the direction $2$ links in the direction $1$
and in the direction $1$ links in the direction $2$ \cite{PAPONE}),
that we call $\Omega(x,L)$, which we plot in fig. \ref{F-OLOL} for
various $L$ ($x\ge 1$).  The different sets of points are plotted
versus $s(x,L)\equiv(1/x+1/(L-x))^{\lambda}$.  If we chose the value
$\lambda=2.1$ the data at given $L$ can be very well fitted by a
linear functions, i.e. the lines in fig. \ref{F-OLOL}:

\begin{equation}
  \Omega(x,L)=a(L) + b(L) \ s(x,L)\ .
  \label{E-AB}
\end{equation}
We plot in fig. \ref{F-AB} $a(L)$ and $b(L)$ from (\ref{E-AB}) and
the best linear fits. If for $L\to\infty$ $1-q_{l}(0)$ would go to
zero, this correlation function should become identically zero and
consequently the two functions, $a(L)$ and $b(L)$, should extrapolate
to zero when $L \to \infty$.  However the slope $b(L)$ does not show
any tendency to go to zero: the fact that it increases (slightly) with
$L$ is a substantial indication that it does not converge to zero in
the infinite volume limit.  The intercept is fitted very well by a
$1/L$ behavior without corrections and it extrapolates to a non zero
value.  Moreover the correlation function at large distances should be
the square of the expectation values and consequently we have the
relation $a(L)|_{L=\infty}=(1-q_{l}( 0))^{2}$.  This relation gives a
value of $1-q_l( 0)=0.26\pm 0.02$, that compares very nicely to the
values we have discussed earlier.

Let us start summarizing our results. We have applied a bulk
perturbation, and shown that the results substantiate in a perfect way
the ones obtained by changing the boundary conditions
\cite{PAPONE}. The quantity $1 -q_l( 0)$ has a non-zero value
that compares very well to the one determined in \cite{PAPONE}. We
have also determined that $\overline{q_l}$ behaves very similarly, and
we have estimated its value in the infinite volume limit (four
different estimates, affected by different systematic errors are $.24
\pm .01$, $.26 \pm .02$, $.30 \pm .01$, $.33 \pm .02$, the errors
being purely statistical). We have shown that very plausibly in the
limit $L\to\infty$ $q_l$ is a well behaved function of $q^2$, in
strict analogy to what happens in the mean field RSB theory. We have
analyzed site overlap and link overlap spatial correlation functions,
and we have shown they perfectly obey scaling laws implied by a RSB
scenario. Again, this results is quantitative, in which it leads to an
estimate of $q_l( 0)$ coherent with the ones given before.

The RSB scenario is perfectly compatible with the whole set of data.
These results are consistent with the recent detection of large scale
excitations in spin glass ground states \cite{MARTIN,PALYOU2}.  The
behavior we find connects smoothly with the results obtained by
simulations at finite temperature \cite{MPRRZ} (e.g. the
overlap-overlap correlation decays as the distance to a power
$\phi=.4$, which is not far from the value obtained from simulation at
finite temperature).  We think that we have conclusively shown that
the ground state structure of $3D$ Ising spin glasses with Gaussian
quenched random couplings is not trivial.

EM thanks for the very nice hospitality the SPhT of Saclay CEA and the
LPTMS of Universit\'e Paris Sud, Orsay, where part of this work has
been done.  We are grateful to O. Martin, M. Palassini and P. Young for
useful correspondences and conversations.

\end{multicols}
\end{document}